\documentclass[preprint,prb]{revtex4}
\usepackage{amsmath}
\usepackage{amssymb}
\usepackage{amsthm}

\newcommand{\gradv}{\boldsymbol{\nabla}}
\def\v#1{{\bf#1}}

\begin{document}

\title{The covariant formulation of Maxwell's equations\\ expressed in a form independent of specific units}
\author{Jos\'e A Heras$^{1,2}$}
\email{herasgomez@gmail.com}
\author{G B\'aez$^{1}$}
\email{gbaez@correo.azc.uam.mx}
\affiliation{$^1$Departamento de Ciencias B\'asicas, Universidad Aut\'onoma Metropolitana Unidad Azcapotzalco, Av. San Pablo No. 180, Col. Reynosa, 02200, M\'exico D. F. M\'exico\\
$^2$Departamento de F\'isica y Matem\'aticas, Universidad Iberoamericana, Prolongaci\'on Paseo de la Reforma 880, M\'exico D. F. 01210, M\'exico}

\begin{abstract}
The covariant formulation of Maxwell's equations can be expressed in a form independent of the usual systems of units by introducing 
the constants $\alpha,\beta$ and $\gamma$ into these equations. Maxwell's equations involving these constants are then specialized to the most commonly used systems of units: Gaussian, SI and Heaviside-Lorentz by giving the constants $\alpha,\beta$ and $\gamma$ the values appropriate to each system.
\end{abstract}

\maketitle
\noindent{\bf 1. Introduction}\\
\noindent The idea of expressing Maxwell's equations in a form independent of specific units has been recurrent in the literature [1-9]. A common strategy to accomplish this idea consists in introducing  a number of unspecified constants into Maxwell's equations. These equations are then specialized to any desired system by giving the constants the values appropriate to that system. For example, Gelman [4] introduced five constants into Maxwell's equations which specialize to obtain 
these equations in Gaussian, International System (SI), Heaviside-Lorentz (HL), electrostatic (esu) and electromagnetic (emu) units. Similarly, Jackson [5] introduced four constants into Maxwell's equations  which properly specialize to yield these equations in the above mentioned units. Usual textbooks on electromagnetism do not generally employ neither esu nor emu units. Almost all undergraduate-level textbooks employ SI units and some of the widely adopted graduate-level textbooks employ Gaussian units. Thus we can ignore the esu and emu units. In this case three constants are sufficient to express Maxwell's equations in a way independent of units like Gaussian and SI units. We have done this in a recent paper [9] in which we have introduced the constants $\alpha,\beta$ and $\gamma$ into Maxwell's equations. In terms of these constants, Maxwell's equations can subsequently be specialized to the most commonly used systems of units: Gaussian, SI and HL. 

In the $\alpha\beta\gamma$-system, Maxwell's equations for sources in vacuum read [9]:   
\begin{align}
\gradv\cdot\v E&=\alpha\rho,\\ 
\gradv\cdot\v B&= 0,\\
\gradv\times \v E+\gamma\frac{\partial \v B}{\partial t}&= 0,\\
\gradv\times \v B-\frac{\beta}{\alpha}\frac{\partial \v E}{\partial t}
&=\beta\v J.
\end{align}
In Table I we display the values of $\alpha,\beta$ and $\gamma$ corresponding to Gaussian, SI and HL units. 

\begin{table}[h]
\begin{center}
\begin{tabular}{|l|l|l|l|}
\hline
\;System & $\;\;\alpha$ & $\;\;\:\beta$ & $\;\;\gamma$\\
\hline
\;Gaussian & $\;\;4\pi$ & $\:4\pi/c $ & $\;1/c$\\ \hline
\;SI & $1/\epsilon_0$&$\;\;\mu_0$& \;\;1\\ \hline
\;Heaviside-Lorentz & $\;\;1$ & $\;\;1/c$ & $\;1/c$\\ 
\hline
\end{tabular}
\end{center}
\caption{\label{tab1} The $\alpha\beta\gamma$-system}
\end{table}

\noindent The constants $\alpha, \beta$ and $\gamma$ satisfy the relation 
\begin{align}
\alpha=\beta\gamma c^2,
\end{align}
 being $c$ the speed of light in a vacuum. For example, if we insert $\alpha=1/\epsilon_0,\beta=\mu_0$ and $\gamma=1$ into equations (1)-(4) then  we obtain Maxwell's equations in SI units. As noted by Gelman [4], the constants $\alpha$, $\beta$ and $\gamma$
`... are experimental constants in the sense that they occur in the experimental-vacuum force laws.' Table I is actually a reduced version of the table given by Gelman [4], which includes five constants.  In an abuse of language and for simplicity, let us say that equations (1)-(4) are Maxwell's equations in 
``$\alpha\beta\gamma$ units," on the understanding that $\alpha,\beta$ and $\gamma$ are not themselves units but rather constants representing units.

Let us write some well-known electromagnetic 
quantities in $\alpha\beta\gamma$ units. 
\vskip 10pt
\noindent Lorentz force:
\begin{align}
\v F= q(\v E +\gamma\v v\times\v B).
\end{align}
Wave equations for the electric and magnetic fields:
\begin{align}
\Box^2 \v E=\alpha\gradv\rho+ \beta\gamma\frac{\partial \v J}{\partial t}\quad{\rm and}\quad \Box^2\v B=-\beta\gradv\times \v J,
\end{align}

where $\Box^2=\nabla^2-(1/c^2)\partial^2/\partial t^2.$
\vskip 10pt
\noindent Time-dependent extensions of the Coulomb and Biot-Savart laws [9]: 
\begin{align}
 \v E &=\frac{\alpha}{4\pi}\int d^3x'\bigg(\frac{\hat{\v R}}{R^2}[\rho]+\frac{\hat{\v R}}{Rc}\left[\frac{\partial \rho}{\partial t}\right]
-\frac{1}{Rc^2}\left[\frac{\partial \v J}{\partial t}\right]\bigg),\\
 \v B&=\frac{\beta}{4\pi}\int d^3x' \bigg([\v J]\times\frac{\hat{\v R}}{R^2 }+\bigg[\frac{\partial \v J}{\partial t}\bigg]\times\frac{\hat{\v R}}{R c}\bigg),
\end{align}
Coulomb and Biot-Savart laws:
\begin{align}
 \v E &=\frac{\alpha}{4\pi}\int d^3x'\frac{\hat{\v R}\rho}{R^2}\quad {\rm and}\quad \v B=\frac{\beta}{4\pi}\int d^3x'\frac{ \v J\times\hat{\v R}}{R^2 }.
\end{align}
\noindent Energy density:
\begin{align}
u= \frac{\v E^2/\alpha+ \gamma\v B^2/\beta}{2}.
\end{align}
Poynting vector:
\begin{align}
\v S=\frac{1}{\beta}\v E\times \v B.
\end{align}
Electromagnetic momentum:
\begin{align}
\v g= \frac{\gamma}{\alpha}\v E\times \v B.
\end{align}
Electric field in terms of potentials:
\begin{align}
\v E=-\gradv\Phi-\gamma\frac{\partial\v A}{\partial t}.
\end{align}
Maxwell's equations in terms of potentials:
\begin{align}
\nabla^2\Phi+\gamma\frac{\partial}{\partial t}\gradv\cdot{\v A}&= -\alpha\rho,\\
\Box^2{\v A}- \gradv\bigg(\gradv\cdot\v A+\frac{\beta}{\alpha}\frac{\partial\Phi}{\partial t}\bigg)&=-\beta\v J.
\end{align}
Lorenz condition:
\begin{align}
\gradv\cdot\v A+\frac{\beta}{\alpha}\frac{\partial\Phi}{\partial t}=0.
\end{align}
Wave equations for potentials in the Lorenz gauge:
\begin{align}
\Box^2{\Phi}= -\alpha\rho\quad{\rm and}\quad\Box^2{\v A}=-\beta\v J.
\end{align}
Gauge transformation of the scalar potential:
\begin{align}
\Phi'=\Phi-\gamma\frac{\partial\Lambda}{\partial t}.
\end{align}

In material media described by the polarization vector $\v P$ and the magnetization vector $\v M$, equations (1) and (4) properly generalize to
 \begin{align}
\gradv\cdot\v E&=\alpha(\rho-\gradv\cdot\v P),\\ 
\gradv\times \v B-\frac{\beta}{\alpha}\frac{\partial \v E}{\partial t}
&=\beta\bigg(\v J +\frac{1}{\gamma}\gradv\times \v M+\frac{\partial \v P}{\partial t}\bigg).
\end{align}
However, it is traditional to formulate these equations in terms of the well-known  vectors $\v D$ and $\v H$. We will see in the Appendix A that such a formulation demands the introduction of two additional constants [4].

The idea of writing Maxwell's equations in a way independent of units like SI or Gaussian units has been mostly developed for the familiar three-dimensional form of these equations but not for its covariant four-dimensional formulation. Standard undergraduate texts like 
Griffiths's book [10] employs SI units in the covariant formulation of Maxwell equations. However, graduate texts like Jackson's book [11] employs Gaussian units in such a formulation. It is worthwhile then to write the covariant Maxwell equations in $\alpha\beta\gamma$ units in order to achieve that these  equations turn out to be independent of the familiar SI and Gaussian units. Furthermore, Maxwell's equations extended to include magnetic monopoles seem not to have been written, as far as we know, in a form independent of specific units like SI or Gaussian units. Therefore, it would be also useful to express Maxwell's equations with magnetic monopoles in  $\alpha\beta\gamma$ units.

The plan of this paper is as follows. In section 2 we express the covariant form of Maxwell's equations in $\alpha\beta\gamma$ units and write some important electromagnetic quantities in such units. In section 3 we discuss some pedagogical advantages of the
$\alpha\beta\gamma$-system. In section 4 we present our conclusions. In Appendix A we discuss the idea of writing equations (20) and (21) in terms of the vectors $\v D$ and $\v H$. Finally, in Appendix B we express both the vector form of Maxwell's equations with magnetic monopoles and the covariant formulation of these equations in $\alpha\beta\gamma$ units.\\

\noindent{\bf 2. Covariant formulation of Maxwell's equations}\\ 
 \noindent Greek indices $\mu, \nu, \kappa ...$ run from 0 to 3; Latin indices $i,j,k,...$ run from 1 to 3;
$x=x^{\mu}=(x^0,x^i)=(ct,\v x)$ is the field point and $x'=x'^{\mu}=(x'^0,x'^i)=(ct',\v x')$ the source point; 
the signature of the metric $\eta^{\mu\nu}$ of the Minkowski spacetime  is $(+,-,-,-);$ $\varepsilon^{\mu\nu\alpha\beta}$ is the 
totally antisymmetric four-dimensional tensor with $\varepsilon^{0123}=1$ and $\varepsilon^{ijk}$ is the totally antisymmetric three-dimensional tensor with 
$\varepsilon^{123} = 1.$ Summation convention on repeated indices is adopted. A four-vector in spacetime can be represented in the so-called (1+3) notation as $F_\nu=(f_0,\v F),$ where $f_0$ is its time component and $\v F$ its space component [12].
Derivatives in spacetime are defined by $\partial_\mu=[(1/c)\partial/\partial t, \gradv]$ and  $\partial^\mu=[(1/c)\partial/\partial t,- \gradv].$

The source of the electromagnetic field tensor $F^{\mu\nu}$ is the four-current 
\begin{align}
J^\mu=(c\rho,\v J).
\end{align}
The tensor $F^{\mu\nu}$ satisfies the Maxwell equations in $\alpha\beta\gamma$ units:
 \begin{align}
\partial_\mu F^{\mu\nu} &=  \beta J^\nu,\\
\partial_\mu\!{^*}\!F^{\mu\nu} &=  0,
\end{align}
where $^*\!F^{\mu\nu}=(1/2)\varepsilon^{\mu\nu\kappa\sigma}F_{\kappa\sigma}$ is the dual of $F^{\mu\nu}$. The tensor $F^{\mu\nu}$ is defined by its components
\begin{equation}
F^{i0}=\frac{\beta c}{\alpha}(\v E)^i \quad{\rm and}\quad F^{ij}=-\varepsilon^{ijk}(\v B)_k.
\end{equation}
where $(\v E)^i$ and $(\v B)_k$ represent the components of the electric and magnetic fields. We call $F^{i0}$ and $F^{ij}$ the polar and axial components  of $F^{\mu\nu}$  respectively [12]. We note that only the polar component involves constants of the $\alpha\beta\gamma$-system. 
The components of the dual tensor $^*\!F^{\mu\nu}$ can be obtained from those of $F^{\mu\nu}$ by making the following dual changes: 
\begin{equation}
\frac{\beta c}{\alpha}(\v E)^i\rightarrow (\v B)^i\quad{\rm and}\quad (\v B)_k \rightarrow -\frac{1}{\gamma c}(\v E)_k. 
\end{equation}
Notice that $\beta c/\alpha=1/(\gamma c).$ Therefore
\begin{equation}
^*\!F^{i0}=(\v B)^i\quad{\rm and}\quad ^*\!F^{ij}=\frac{1}{\gamma c }\varepsilon^{ijk}(\v E)_k.
\end{equation}
Only the axial component of $^*\!F^{\mu\nu}$ involves constants of the $\alpha\beta\gamma$-system. With the aid of the above definitions, we can write the following four-vectors in the (1+3) notation:
\begin{align}
\partial_{\mu}F^{\mu\nu}=&\bigg (\frac{\beta c}{\alpha}\nabla \cdot \v E,\;\nabla \times \v B-\frac{\beta}{\alpha}\frac{\partial \v E}{\partial t} \bigg ),\\
\partial_{\mu}\!{^*}\!F^{\mu\nu}=&\bigg (\nabla \cdot \v B, \; -\frac{1}{\gamma c}\nabla \times \v E-\frac 1c\frac{\partial \v B}{\partial t} \bigg) .
\end{align}
Let us emphasize the pedagogical importance of the four-vectors (28) and (29). They have previously  been considered in specific units 
[12,13]. Equations (28) and (29) together with equation (22) make transparent the derivation of the familiar three-dimensional formulation of Maxwell's equations from its covariant four-dimensional formulation. In fact, to obtain the vector form of Maxwell's equations from equations (23) and (24), we first make equal the time and space components in both sides of equation (23) and use equations (22) and (28). As a result 
we obtain equations (1) and (4). Next we make zero the time and space components in equation (24) and use equation (29) to obtain the remaining equations (2) and (3).

The four-vector (28) can be obtained as follows
\begin{align}
\partial _{\mu}F^{\mu\nu}=&\bigg (\partial_{\mu}F^{\mu 0},\; \partial_{\mu}F^{\mu j}\bigg)\nonumber \\
=& \bigg (\partial_{i}F^{i0}, \;
\partial _{0}F^{0 j}+\partial_{i}F^{ij} \bigg)\nonumber\\
=&\bigg (\frac{\beta c}{\alpha}\partial_i(\v E)^i, -\frac{\beta}{\alpha}\frac{\partial}{\partial t}(\v E)^j\! +\!\varepsilon^{jik}\partial_{i}(\v B)_k \bigg),
\end{align}
where we have used equations (25). Clearly, equation (30) is the same as equation (28). By a similar calculation we can obtain the four-vector (29). 

Let us specialize to Gaussian units. In this case equations (23) and (24) take the form
\begin{equation}
\partial_\mu F^{\mu\nu} = \frac{4\pi}{c}J^\nu \quad{ \rm and}\quad 
\partial_\mu\!{^*}\!F^{\mu\nu} =  0,
\end{equation}
where the polar and axial components of $F^{\mu\nu}$ are 
$F^{i0}=(\v E)^i$ and $F^{ij}=-\varepsilon^{ijk}(\v B)_k,$
and therefore 
$^*\!F^{i0}=(\v B)^i$ and $^*\!F^{ij}=\varepsilon^{ijk}(\v E)_k.$
Equations (28) and (29) become
\begin{align}
\partial _{\mu}F^{\mu\nu}=&\bigg (\nabla \cdot \v E,\;\nabla \times \v B-\frac{1}{c}\frac{\partial \v E}{\partial t} \bigg ),\\
\partial _{\mu}\!{^*}\!F^{\mu\nu}=&\bigg (\nabla \cdot \v B, \; -\nabla \times \v E-\frac 1c\frac{\partial \v B}{\partial t} \bigg) .
\end{align}
From equations (22), (31)-(33) we obtain the familiar Maxwell's equations in Gaussian units. 
 
 We consider now SI units. In this case equations (23) and (24) read
\begin{equation}
\partial_\mu F^{\mu\nu} = \mu_0J^\nu \quad{\rm and}\quad
\partial_\mu\!{^*}\!F^{\mu\nu} =  0,
\end{equation}
where the polar and axial components of  $F^{\mu\nu}$ are $
F^{i0}=(\v E)^i/c$ and $F^{ij}=-\varepsilon^{ijk}(\v B)_k.$ It follows that
$^*\!F^{i0}=(\v B)^i$ and $^*\!F^{ij}=\varepsilon^{ijk}(\v E)_k/c.$
Therefore equations (28) and (29) become
\begin{align}
\partial _{\mu}F^{\mu\nu}=&\bigg (\frac{1}{c}\nabla \cdot \v E,\;\nabla \times \v B-\epsilon_0\mu_0\frac{\partial \v E}{\partial t} \bigg ),\\
\partial _{\mu}\!{^*}\!F^{\mu\nu}=&\bigg (\nabla \cdot \v B, \; -\nabla \times \v E-\frac{\partial \v B}{\partial t} \bigg),
\end{align}
where we have used $\epsilon_0\mu_0=1/c^2$. From equations (22), (34)-(36), we obtain the three-dimensional form of Maxwell's equations in SI units.

Finally, in the HL units equations (23) and (24) read
\begin{equation}
\partial_\mu F^{\mu\nu} = \frac{1}{c}J^\nu \quad {\rm and}\quad
\partial_\mu\!{^*}\!F^{\mu\nu} = 0,
\end{equation}
where $F^{i0}=(\v E)^i$ and $F^{ij}=-\varepsilon^{ijk}(\v B)_k.$ These components
imply 
$^*\!F^{i0}=(\v B)^i$ and $ ^*\!F^{ij}=\varepsilon^{ijk}(\v E)_k.$
In this case equations (28) and (29) become
\begin{align}
\partial _{\mu}F^{\mu\nu}=&\bigg (\nabla \cdot \v E,\;\nabla \times \v B-\frac{1}{c}\frac{\partial \v E}{\partial t} \bigg ),\\
\partial _{\mu}\!{^*}\!F^{\mu\nu}=&\bigg (\nabla \cdot \v B, \; -\nabla \times \v E-\frac 1c\frac{\partial \v B}{\partial t} \bigg) .
\end{align}
Equations (22), (37)-(39) yield the vector form of Maxwell's equations in HL units. 

To finish this section let us write some covariant 
quantities in $\alpha\beta\gamma$ units. 
\vskip 10pt
\noindent Lorentz force:
\begin{align}
\frac{dP^\mu}{d\tau}= \gamma F^{\mu\nu}J_{\nu}.
\end{align}
Wave equations for the electromagnetic field:
\begin{align}
\partial_\sigma\partial^\sigma F^{\mu\nu}=\beta(\partial^\mu J^\nu-\partial^\nu J^\mu).
\end{align}
Retarded solution of Maxwell's equations:
\begin{align}
F^{\mu\nu}(x)=\beta\int d^4 x'G(x,x')(\partial{'^\mu} J^\nu(x')-\partial'{^\nu J^\mu(x')}),
\end{align}
where $G(x,x')$ satisfies $\partial_\sigma \partial^\sigma G=\delta(x-x')$ with $\delta(x-x')$ being the four-dimensional Dirac delta function.
\vskip 10pt
\noindent The four-potential in the $(1+3)$ notation:
\begin{align}
A^{\mu}=\bigg(\frac{\beta c}{\alpha}\Phi, \v A\bigg).
\end{align}
\noindent Notice that the Lorenz condition $\partial_\mu A^\mu=0$ yields the three-dimensional form of this condition expressed in equation (17). Notice also that 
only the temporal component of $A^\mu$ involves constants of the $\alpha\beta\gamma$-system. 
\vskip 10pt
\noindent Inhomogeneous Maxwell equations in terms of the four-potential:
\begin{align}
\partial_\sigma\partial^\sigma A^{\mu}-\partial^\mu\partial_\nu A^{\nu}=\beta J^\nu.
\end{align}
Wave equation for the four-potential in the Lorenz gauge:
\begin{align}
\partial_\sigma\partial^\sigma A^{\nu}=\beta J^\nu.
\end{align}
Lagrangian for the electromagnetic field:
\begin{align}
{\cal L}= -\frac{\gamma}{4\beta}F_{\mu\nu}F^{\mu\nu}-\gamma J_\mu A^\mu.
\end{align}

\noindent Symmetric stress tensor:
\begin{align}
\Theta^{\mu\nu}= -\frac{\gamma}{\beta}\bigg(F{^\mu}{_\sigma} F^{\sigma\nu} + \frac{1}{4}\eta^{\mu\nu}F_{\sigma\theta}F^{\sigma\theta} \Bigg).
\end{align}

It is sometimes claimed that the covariant formulation of electrodynamics is simpler and more elegant in Gaussian units than in SI units. A simple reading of the formulas given in this section shows that the covariant formulation of electrodynamics in $\alpha\beta\gamma$-units is as simple and elegant as in Gaussian units.\\

\noindent{\bf 3. Discussion} \\
The first two editions [5,14] of Jackson's {\it classical electrodynamics} employed Gaussian units. This authoritative  book strongly endorsed 
the tradition of teaching advanced electrodynamics in Gaussian units. However, most undergraduate textbooks on electromagnetism employ now SI units. 
Some students could not be comfortable with this situation since they study undergraduate electrodynamics in SI units and graduate electrodynamics in Gaussian units. In the third edition [10] of his book, Jackson has proposed a Solomonic solution to this problem. He has introduced SI units, which seem more situed for practical applications than Gaussian units, into the first ten chapters and retained Gaussian units for the
remainder of the book since such units seem more suited to relativistic electrodynamics than SI units. Thus graduate students actually work with equations in both SI and Gaussian units, which does not seem to be the best alternative from a pedagogical point of view. In this context, 
the $\alpha\beta\gamma$-system may be a pedagogical alternative for undergraduate and graduate students who can solve electromagnetic problems without having to work in a specific system of units. At the end of their calculations, they can present their results in any desired system by making use of the Table I. \\

\noindent {\bf 4. Conclusions}\\
In this paper we have reconsidered the old idea [1-5] of writing electromagnetic equations in a form independent of specific 
units like SI or Gaussian units. We have followed the work of Gelman [4] who introduced five constants in Maxwell's equations 
which properly specialize to obtain these equations in Gaussian, SI, HL, esu and emu units. Because the last two units are no longer in use in textbooks, we have adopted here only three constants ($\alpha, \beta$ and $\gamma$) to include Gaussian, SI and  HL units [9]. In particular we have expressed the covariant Maxwell equations in the $\alpha\beta\gamma$-system. Then we have specialized these equations to Gaussian, SI and HL systems of units by giving the constants $\alpha, \beta$ and $\gamma$ the values appropriate to these systems (Table I). 
The covariant formulation of electrodynamics in $\alpha\beta\gamma$-units is simple and elegant. For completeness we have also expressed   in the $\alpha\beta\gamma$-system  both the vector form of Maxwell equations with magnetic monopoles and the covariant form of these equations.

\vskip 15pt
\noindent {\bf Acknowledgment} 

\noindent The support of the Fondo UIA-FICSAC is gratefully acknowledged.

\appendix 

\section{}\noindent{\bf Maxwell's equations with the vectors $\v D$ and $\v H$}\\

\noindent We can express equations (20) and (21) as follows  
\begin{align}
\gradv\cdot\v D=\alpha\rho,\\
\gradv\times \v H-\frac{\beta}{\alpha}\frac{\partial \v D}{\partial t}=\beta\v J,
\end{align}
where the vectors $\v D$ and $\v H$ are defined as 
\begin{align}
\v D=\v E+\alpha \v P\quad {\rm and}\quad \v H=\v B-\frac{\beta}{\gamma}\v M.
\end{align}
In Gaussian units, equations (A3) take their usual form: $\v D=\v E+4\pi\v P$  and $\v H= \v B-4\pi\v M.$ But in SI units, equations (A3) take the unusual form: $\v D=\v E+\v P/\epsilon_0$  and $\v H=\v B-\mu_0\v M.$
As is well known, the usual definition of the vectors $\v D$ and $\v H$ in SI units is given by $\v D=\epsilon_0\v E+\v P$  and $\v H=(1/\mu_0)\v B-\v M.$ Therefore equations (A1)-(A3) are not consistent in SI units. Alternatively, we can also write equations (20) and (21) as 
\begin{align}
\gradv\cdot\v D=\rho,\\
\gradv\times \v H-\frac{\partial \v D}{\partial t}=\v J.
\end{align}
whenever the vectors $\v D$ and $\v H$ are defined as 
\begin{align}
\v D=\frac{1}{\alpha}\v E+\v P\quad {\rm and}\quad \v H=\frac{1}{\beta}\v B-\frac{1}{\gamma}\v M.
\end{align}
In Gaussian units equations (A6), take the unusual form: $\v D=\v E/(4\pi)+\v P$  and $\v H= c\v B/4\pi
-c\v M.$ But in SI units equations (A6) take the standard form: $\v D=\epsilon_0\v E+\v P$  and $\v H=(1/\mu_0)\v B-\v M.$ Therefore equations (A4)-(A6) are not consistent in Gaussian units. 

The above attempts to consistently write equations (20) and (21) in terms of the vectors $\v D$ and $\v H$ defined either by equations (A3) or by equations (A6) have not been successful. If we insist in keeping unchanged the traditional definitions of $\v D$ and $\v H$ in both Gaussian and SI units then we must appropriately 
rewrite equations (20) and (21). Following Gelman [4], let us multiply equation (20) by the constant $k_1$ and write the resulting equation as 
\begin{align}
\gradv\cdot\v D=k_1\alpha\rho,
\end{align}
where the vector $\v D$ is defined as 
\begin{align}
\v D=k_1(\v E+\alpha \v P).
\end{align}
Next, we multiply equation (21) by the constant $k_2$ to obtain
\begin{align}
\gradv\times k_2\bigg(\v B-\frac{\beta}{\gamma}\v M\bigg)-\frac{\beta k_2}{\alpha}\frac{\partial}{\partial t}(\v E+\alpha\v P)=k_2\beta\v J.
\end{align}
From equation (A8) it follows that  $\v E+\alpha \v P=\v D/k_1$. Using this result into equation (A9) we obtain an equation that can be written as
\begin{align}
\gradv\times \v H-\frac{\beta}{\alpha}\frac{k_2}{k_1}\frac{\partial\v D}{\partial t}=k_2\beta\v J,
\end{align}
 where the vector $\v H$ is defined as 
\begin{align}
\v H=k_2\bigg(\v B-\frac{\beta}{\gamma}\v M\bigg).
\end{align}
For suitable values  of the constants $k_1$ and  $k_2$, equations (A7), (A8), (A10) and (A11) yield the corrects forms for Gaussian, SI and HL units. 

In Gaussian units and taking the values $k_1=1$ and $k_2=1$ it follows that equations (A7) and (A10) take their usual form in Gaussian units: 
\begin{align}
\gradv\cdot\v D=4\pi\rho,\\
\gradv\times \v H-\frac{1}{c}\frac{\partial \v D}{\partial t}=\frac{4\pi}{c}\v J,
\end{align}
where $\v D=\v E+4\pi\v P$  and $\v H= \v B-4\pi\v M$ follow from equations (A8) and (A11).

In SI units and making $k_1=\epsilon_0$ and $k_2=1/\mu_0$ we see that equations (A7) and (A10) yield Maxwell's source equations in material media expressed in SI units
\begin{align}
\gradv\cdot\v D=\rho,\\
\gradv\times \v H-\frac{\partial \v D}{\partial t}=\v J,
\end{align}
where $\v D=\epsilon_0\v E+\v P$  and $\v H=(1/\mu_0)\v B-\v M$ are obtained from equations (A8) and (A11).

Finally, in HL units and assuming values $k_1=1$ and $k_2=1,$ equations (A7) and (A10) reproduce Maxwell's source equations in material media expressed
in HL units: 
\begin{align}
\gradv\cdot\v D=\rho,\\
\gradv\times \v H-\frac{1}{c}\frac{\partial \v D}{\partial t}=\frac{1}{c}\v J,
\end{align}
where $\v D=\v E+\v P$  and $\v H= \v B-\v M$ are derived from equations (A8) and (A11).

By assuming the specified values for $k_1$ and $k_2$ in each one of the above considered systems of units, equations (A7) and (A10) represent consistently the inhomogeneous Maxwell equations in terms of $\v D$ and $\v  H$ whenever such vectors are defined by equations (A8) and (A11). Table II extends Table I to include the specific values of $k_1$ and $k_2$ for Gaussian, SI and HL units. Table II is also a reduced form of the table previously given by Gelman [4].

\begin{table}[h]
\begin{center}
\begin{tabular}{|l|l|l|l|l|l|}
\hline
\;System & $\;\;\alpha$ & $\;\;\:\beta$ & $\;\;\gamma$&$\;k_1$&$\;\;k_2$\\
\hline
\;Gaussian & $\;\;4\pi$ & $\:4\pi/c $ & $\;1/c$& $\;\;1$& $\;\;\;1$\\ \hline
\;SI & $1/\epsilon_0$&$\;\;\mu_0$& \;\;1& $\;\epsilon_0$& $1/\mu_0$\\ \hline
\;Heaviside-Lorentz & $\;\;1$ & $\;\;1/c$ & $\;1/c$&$\;\;1\;$& $\;\;\;1$\\ 
\hline
\end{tabular}
\end{center}
\caption{\label{tab2} Extended $\alpha\beta\gamma$-system}
\end{table}

We note that the introduction of the constants $k_1$ and $k_2$ is a matter of convenience. Such constants are not connected with experimental aspects like $\alpha,\beta$ and $\gamma$. In other words: if one disregards the idea of formulating equations (20) and (21) in terms of $\v D$ and $\v H$ in such a way that the re-formulated equations yield the conventional definitions of these vectors in both Gaussian and SI units, then the requirement of introducing $k_1$ and $k_2$ turns out to be unnecessary. After all, equations (20) and (21) as they stand unambiguously describe in  $\alpha\beta\gamma$ units the electric and magnetic fields in material media. 
\section{}\noindent{\bf Maxwell's equations with magnetic monopoles}
\vskip5pt
\noindent In $\alpha\beta\gamma$ units  the Maxwell equations with magnetic monopoles can be written as:
\begin{align}
\gradv\cdot\v E&=\alpha\rho_e,\\ 
\gradv\cdot\v B&= \frac{\beta}{\gamma}\rho_m,\\
\gradv\times \v E+\gamma\frac{\partial \v B}{\partial t}&= -\beta\v J_m,\\
\gradv\times \v B-\frac{\beta}{\alpha}\frac{\partial \v E}{\partial t}
&=\beta\v J_e,
\end{align}
where $\rho_e$ and $\v J_e$ are the electric charge and current densities and $\rho_m$ and $\v J_m$ the magnetic charge and current densities.
In Gaussian units, equations (B1)-(B4) take the form
\begin{align}
\gradv\cdot\v E&=4\pi\rho_e,\\ 
\gradv\cdot\v B&= 4\pi\rho_m,\\
\gradv\times \v E+\frac{1}{c}\frac{\partial \v B}{\partial t}&= -\frac{4\pi}{c}\v J_m,\\
\gradv\times \v B-\frac{1}{c}\frac{\partial \v E}{\partial t}
&=\frac{4\pi}{c}\v J_e,
\end{align}

In SI units, equations (B1)-(B4) read
\begin{align}
\gradv\cdot\v E&=\frac{1}{\epsilon_0}\rho_e,\\ 
\gradv\cdot\v B&= \mu_0\rho_m,\\
\gradv\times \v E+\frac{\partial \v B}{\partial t}&= -\mu_0\v J_m,\\
\gradv\times \v B-\epsilon_0\mu_0\frac{\partial \v E}{\partial t}
&=\mu_0\v J_e.
\end{align}
The unit of the magnetic charge in these equations is the Ampere-meter (see comment at the end of this paper).
Equations (B9)-(B12) appear, for example, in the books of Griffiths [15], Portis [16] and Vanderline [17]. 

In HL units, equations (B1)-(B4) are given by 
\begin{align}
\gradv\cdot\v E&=\rho_e,\\ 
\gradv\cdot\v B&= \rho_m,\\
\gradv\times \v E+\frac{1}{c}\frac{\partial \v B}{\partial t}&= -\frac{1}{c}\v J_m,\\
\gradv\times \v B-\frac{1}{c}\frac{\partial \v E}{\partial t}
&=\frac{1}{c}\v J_e.
\end{align}

The covariant form of Maxwell's equations with magnetic monopoles in $\alpha\beta\gamma$ units are 
\begin{align}
\partial_\mu F^{\mu\nu} =  \beta J_e^\nu\quad {\rm and }\quad \partial_\mu\!{^*}\!F^{\mu\nu} = \frac{\beta}{\gamma c}J_m^\nu,
\end{align}
where the conserved electric and magnetic four-currents are defined by
\begin{align}
J_e^\mu=(c\rho_e,\v J_e) \quad{\rm and}\quad  J_m^\mu=(c\rho_m,\v J_m). 
\end{align}
Using equations (28), (29) and  (B18) into equations (B17) we obtain Maxwell's equations with magnetic monopoles 
displayed in equations (B1)-(B4). We leave the reader the problem of extending equations (40)-(47) to include magnetic monopoles.

Comment: a referee has pointed out that the convention adopted in the SI equations (B10) and (B11) that magnetic charges have units of Ampere-meters {\it is not  natural}. The Gauss law $\oint\v D\cdot d\v S=Q_e$
saying that the electric flux over a closed surface is equal to electric charge
included within the surface, has its magnetic analogue: $\oint\v B\cdot d\v S=Q_m$ saying that 
the magnetic flux is equal to the magnetic charge. Therefore the magnetic charge has the same unit as the magnetic flux, that is, the Weber.
In this case the factor $\mu_0$ in (B10) and (B11) is superfluous and thus these equations should respectively be replaced by 
\begin{align}
\gradv\cdot\v B= \rho_m \quad {\rm and}\quad
\gradv\times \v E+\frac{\partial \v B}{\partial t}= -\v J_m.
\end{align}
These equations appear in Jackson's book [18].
The referee has also suggested another result to support that the Weber is the {\it natural} unit for a magnetic charge. 
From the Dirac relation [19]: $eg =2\pi n\hbar,$ between the electric charge $e$ and the magnetic charge $g,$  one can check that ${\rm Coulomb\cdot Weber =Joule\cdot second}$, which is the appropriate unit of the action.

In $\alpha\beta\gamma$ units the Gauss law for magnetic charges reads $\oint\v B\cdot d\v S=(\beta/\gamma)Q_m$.
In particular, for SI units this law becomes $\oint\v B\cdot d\v S=\mu_0 Q_m,$ which says that the magnetic flux is equal to $\mu_0$ times the magnetic charge. Since $\mu_0$ has units of Weber/(Ampere-meter) it follows that the unit of the magnetic charge is Ampere-meter. 

In $\alpha\beta\gamma$ units the electric field of an electric charge $e$ and the magnetic field of a magnetic charge $g$ are given by
\begin{align}
\v E=\frac{\alpha e}{4\pi}\frac{\hat{\v R}}{R^2}\quad{\rm and}\quad \v B=\frac{\beta g}{4\pi\gamma}\frac{\hat{\v R}}{R^2}.
\end{align}
It can be shown that the angular momentum stored in these fields, if $e$ and $g$ are separated by a distance $d$, is given by
\begin{align}
\v L=\frac{\beta}{4\pi}\: eg \:\hat{\v k},
\end{align}
where $\hat{\v k}$ is a unit vector directed along the line joining the charges $e$ and $g$. It follows that the Dirac quantization condition in $\alpha\beta\gamma$ units takes the form
\begin{align}
eg=\frac{2\pi}{\beta}n\hbar.
\end{align}
For SI units $eg=2\pi n\hbar/\mu_0$ [20] and therefore $e\mu_0 g=2\pi n\hbar$, from which one can check that 
\begin{align}
{\rm Coulomb\cdot\frac{Weber}{Ampere\cdot meter}Ampere\cdot meter =Joule\cdot second}, 
\end{align}
which is the appropriate unit of the action. Although the Weber seems to be the natural unit for a magnetic charge, the 
$\alpha\beta\gamma$-system is consistent with the result that the magnetic charge has unit of Ampere-meter.

{}
\end{document}